\documentclass[a4paper,12pt]{article}
\usepackage{soul}
\usepackage[usenames,dvipsnames]{color}
\usepackage{jheppub}
\usepackage{esvect}
\usepackage{amsmath, amssymb, slashed, epsf, color, graphicx, latexsym}
\usepackage{epsfig}
\usepackage{graphics}
\usepackage{natbib}
\usepackage[mathlines]{lineno}

\begin{document}
	
	\title{Holographic Constraints on Generalised Rivlin-Ericksen Fluid}
	\author{Suvankar Dutta,}
	\author{Taniya Mandal,}
	\author{Sanhita Parihar}

	\affiliation{Indian Institute of Science Education and Research Bhopal\\
	Bhopal Bypass, Bhopal 462066, India } 
		\emailAdd{suvankar@iiserb.ac.in}
	\emailAdd{tmandal@iiserb.ac.in}
	\emailAdd{sanhita18@iiserb.ac.in}

	\abstract{The Rivlin-Ericksen model is one of the oldest models in fluid dynamics to describe non-Newtonian properties. The model comes with two independent transports at second order. In this paper, we study the relativistic origin of the Rivlin-Ericksen fluid. Starting from a relativistic Weyl invariant uncharged fluid in $3+1$ dimensions, we reduce it over light-cone directions and obtain a generic non-relativistic uncharged fluid in one lower dimension with all possible second order terms in the constitutive relations. We observe that the Rivlin-Ericksen fluid is a subclass of our generalised non-relativistic system. We also compute the holographic values of all the non-relativistic second order transports and find that three of them satisfy a universal constraint relation.}

\maketitle
\section{Introduction}

Fluid dynamics, one of the oldest and widely studied subjects, has been offering new surprises even today. Few experimental evidences and theoretical results in the recent past have created a lot of interests in the field of relativistic hydrodynamics and opened up new avenues for further investigation with many fundamental questions. One of the interesting results in this direction was the discovery of an anomalous transport coefficient in the context of the AdS/CFT correspondence \cite{Banerjee:2008th,Erdmenger:2008rm}. Later in \cite{Son:2009tf} it has been shown that such term in the constitutive relation (energy momentum tensor and other charge currents) is not only allowed by the symmetry but is also indispensable to preserve the second law of thermodynamics. This was the beginning of a new era in the field of relativistic hydrodynamics. 

Non-relativistic systems, on the other hand, are also interesting mainly because they are expected to be realised in low energy experiments, condensed matter physics. Being an effective description of a nearly equilibrium interacting system, the constitutive relations of a fluid dynamic system (relativistic or non-relativistic) are perturbatively written in derivative expansion of fluid variables\footnote{In the relativistic system consideration of second order terms in conserved currents is essential because the first order formalism is inconsistent with causality \cite{Israel:1976tn,ISRAEL1979341,HISCOCK1983466}. See \cite{Kovtun:2019hdm} for a comparing discussion.}. In order to write the independent derivative terms at every order in perturbative expansion, one has to follow a guiding principle. Such a principle could either be empirical or fundamental. For example, to a certain order in derivative expansion one can include all the terms allowed by the symmetry of the system. \emph{In this paper, we follow a fundamental principle to write down the allowed independent terms systematically in the constitutive relations of a non-relativistic uncharged fluid up to second order in derivative expansion.} It is believed that all the interacting systems are relativistic at the microscopic level. A non-relativistic system is an effective low energy (velocity) description of an underlying relativistic system. Therefore, we start with a generic second order relativistic fluid with all possible independent terms in energy momentum tensors allowed by the symmetries \cite{Loganayagam:2008is}. We then take the \emph{non-relativistic limit} and obtain a \emph{reduced} non-relativistic system with independent second derivative terms in its stress tensors and other conserved currents. Taking the non-relativistic limit is not unique and hence the non-relativistic constitutive relations, in general, depend on how the limit is taken \cite{Julia:1994bs,Hassaine:1999hn,Jensen:2014wha,Geracie:2015xfa,Jensen:2014aia,Jensen:2014ama,Lucas:2014sia,Kaminski:2013gca,Jain:2015jla}. In this paper, we consider \emph{light-cone reduction} \cite{Rangamani:2008gi,Brattan:2010bw,Banerjee:2014mka} to obtain the non-relativistic system. 

Light-cone reduction reduces a relativistic field theory to a non-relativistic field theory in one lower space dimension.  This is because the symmetry algebra of the relativistic theory reduces to corresponding non-relativistic symmetry algebra in one lower space dimensions upon light-cone reduction. For example, the Poincar\'e algebra reduces to Galilean algebra and conformal algebra boils down to Schr\"odinger algebra \cite{PhysRevD.5.377,Duval:1984cj,Duval:2008jg,Duval:2009vt,Rangamani:2009zz}. Therefore, light-cone reduction of relativistic constitutive equations renders the non-relativistic constitutive equations for a fluid in one lower dimension. Light-cone reduction of relativistic fluids\footnote{See \cite{Banerjee:2015hra,Banerjee:2015uta,Banerjee:2016qxf,PhysRevD.97.096018} for \emph{null reduction.} of relativistic fluids} up to first order in derivative expansion has been discussed in \cite{Rangamani:2008gi,Brattan:2010bw,Banerjee:2014mka}. Construction of non-relativistic entropy current was also discussed in \cite{Banerjee:2014mka}. A holographic computation of non-relativistic stress tensor and transports have been reported in \cite{Ross:2009ar,Dutta:2018xtr}. However, all the works so far mainly considered constitutive relations up to first order in derivative expansions. In this paper, we consider a parity-odd uncharged relativistic fluid obeying  Poincar\'e and Weyl symmetry in $3+1$ dimensional flat spacetime and use the light-cone reduction to obtain a non-relativistic fluid in $2+1$ dimensional flat space with generic second order terms in constitutive relations.

The appearance of second and higher derivative terms in the constitutive relations of a non-relativistic fluid is well-known since long. Such fluids are broadly known as \emph{non-Newtonian fluid}. The Newtonian fluid possesses a linear relation between viscous stress and rate of strain tensor. To explain the properties of a more general class of fluid like honey, thick oils, paints etc. requires to relax the linear relation between stress and strain tensor, thus leading to the study of non-Newtonian fluid. One way to include such higher derivative terms in stress tensor is by modifying the constitutive relations as suggested by experiments. The other approach is more fundamental - mainly based on symmetries. This has been first studied by Reiner \cite{Reiner} in $1945$. Later in $1948$, Rivlin generalized Reiner's idea for isotropic fluid assuming that the stress tensor depends only on the velocity gradients and is invariant under a coordinate transformation \cite{Rivlin}. This model, known as Reiner-Rivlin model, explains non-Newtonian properties like the normal stress effect, centripetal pump effect and Merrington effect. The model also predicts that in a simple shear flow, the normal stresses in and perpendicular to the plane of shear are equal, that was in contradiction with observations. To explain the difference between two normal stresses the form of the stress tensor requires to be generalised and was given by Rivlin and Ericksen  \cite{RE} in $1955$. They assumed that the higher derivative terms in the stress tensor of an isotropic and homogeneous fluid depend on the gradient of velocity, acceleration, second acceleration and so on and are also frame invariant. However, this model fails to explain gradual stress relaxation. Later in $1960$, Coleman and Noll generalised the Rivlin-Ericksen model further to explain fluid with gradually fading memory \cite{Colenoll}. This generalised version is widely known as Rivlin-Ericksen (RE) fluid and successfully explains flow of a wide class of fluids. The second order RE fluid comes with three independent transport coefficients - one first order transport $n$ (shear viscosity) and two second order transports: the first and second normal stress coefficient denoted by $\alpha_1$ and $\alpha_2$ respectively. It was shown by Dunn and Fosdick \cite{DFS}, that an equilibrium RE fluid  (assuming specific Helmholtz free energy is minimum at equilibrium) satisfying Claussius-Duhem inequality, has both shear viscosity $n$ and the first normal coefficient $\alpha_1$ positive and two second order transports add up to zero : $n,\alpha_1>0$, $\alpha_1+\alpha_2=0$. It has been further studied by Rajagopal and Fosdick \cite{RajaFS}, that shear viscosity and the first normal coefficient are always positive but the sum of the two second order coefficients may not be zero always.

In this paper, we explore the relativistic origin of Rivlin-Ericksen terms in the stress tensor of a non-Newtonian fluid. Starting with a generic second order relativistic parity-odd uncharged fluid in $3+1$ dimensions, we reduce its constitutive relations over the light-cone and obtain a non-relativistic parity-odd uncharged fluid in $2+1$ dimensions up to second order in derivative expansion. At first order the reduced fluid has two independent transports - shear viscosity and thermal conductivity \cite{Rangamani:2008gi,Brattan:2010bw,Banerjee:2014mka}. We observe that at second order  there are total eight independent terms in the stress tensor with four independent transports : $\tilde\xi_\sigma, \tilde\xi_\omega, \tilde \tau_\omega$ and $\tilde \tau_\pi$. Among these, the terms proportional to $\tilde\xi_\sigma$ and $\tilde\tau_\pi$ are present in the RE model. The RE fluid, therefore, is a subset of the generic second order fluid that we have obtained in this paper. The relativistic source of the first RE term ($\tilde \xi_\sigma$) is the square of shear stress tensor. The origin of the second RE transport is interesting - it is proportional to the relaxation time of the relativistic fluid. Apart from the regular RE terms, we also obtain terms dependent on vorticity and temperature gradient at the second order. 

We further compute the holographic values of the second order transports using the \emph{fluid/gravity} correspondence. We use the dictionary between relativistic and non-relativistic transports (table \ref{table}) and use the holographic values of the relativistic ones to find the corresponding non-relativistic values. We find that all four second order transports are not independent. Three of them satisfy a universal relation, namely, $\tilde{\xi}_\sigma+\tilde{\tau}_\omega+\tilde{\tau}_\pi=0$. This relation is holographic, generalisation of \cite{DFS} and consistent with the observation of \cite{RajaFS}.

The plan of the paper is as follows: In section \ref{rel-fluid}, we present the energy-momentum tensor of a relativistic second order fluid. A short discussion about non-relativistic fluid is given in section \ref{nonrel-fluid}. In section \ref{L-C-R}, we discuss light-cone reduction of relativistic fluid and explicitly compute the constitutive relations and transports of non-relativistic fluid in terms of relativistic data up to second order in derivative expansion. We compare our non-relativistic stress tensor with Rivlin-Ericksen fluid and find out the holographic values of non-relativistic transports. Finally, we end our paper with concluding remarks and possible future directions in section \ref{Diss}. 


\section{Second order relativistic fluid}\label{rel-fluid}

A relativistic fluid in $3+1$ dimensions without any chemical potential (uncharged fluid) is specified by \emph{six} parameters : $3$ velocity components (assuming $u^\mu u_\mu=-1$) and three local thermodynamic quantities - energy $E(x)$, temperature $T(x)$ and pressure $P(x)$. However, local thermal equilibrium implies that the thermodynamic variables energy, temperature and pressure satisfy the first law of thermodynamics at every spacetime point. Additionally, $E(x)$, $P(x)$ and $T(x)$ also follow equation of state and Euler relation $E+P= TS$, where $S$ is local entropy density. As a result, a relativistic fluid is specified by three independent components of velocity and one thermodynamic variable, which we choose to be local temperature $T(x)$. The energy momentum tensor of a relativistic fluid therefore depends on temperature and fluid velocities. To find an expression for energy momentum tensor we use its conservation equation
\begin{equation}\label{rel:const}
  \partial_\mu T^{\mu\nu} =0.
  \end{equation}
In $3+1$ dimensions, these are total 4 equations and we also have 4 unknown variables. Hence the system is solvable. Since a fluid is considered to be in local thermal equilibrium and the length scale of variation of fluid variables is much much larger than the mean free path of the system. Hence we express energy momentum tensor perturbatively in derivatives of fluid variables. At every order in derivative expansion, the energy momentum tensor contains independent symmetric second rank tensors constructed out of derivatives of fluid variables. If a fluid obeys any extra symmetry, we include that information in the energy momentum tensor as well. For example, for a conformal fluid, the energy momentum tensor is traceless. Therefore while writing the independent terms at different orders we take care of this constraint. 

Relativistic hydrodynamics has been studied extensively in the last two decades in the context of the fluid/gravity correspondence and also independently. The energy momentum tensor of boundary gauge theory in flat spacetime has been constructed holographically by \cite{Bhattacharyya:2008jc, Baier:2007ix} up to second order in derivative expansion by solving Einstein's equation perturbatively. The second order corrected energy momentum tensor and conserved current of a charged fluid have been constructed in \cite{Banerjee:2008th,Erdmenger:2008rm}. The authors in \cite{Loganayagam:2008is} have given the most generic form of a second order energy momentum tensor and entropy current of a relativistic, conformal fluid in a Weyl covariant manner. The corresponding stress tensor contains all possible independent second rank symmetric terms (up to second order) that are Weyl covariant, transverse and traceless. The transport coefficients are also constrained due to the second law of thermodynamics\footnote{See \cite{Bhattacharyya:2012nq} also.}.

Following \cite{Loganayagam:2008is}, the stress tensor of $(3+1)$ dimensional, uncharged, conformal fluid in a flat spacetime up to second order is given by
  \begin{eqnarray}\label{secondstress}
  T^{\mu\nu} = \left(E+P\right) u^\mu u^\nu +P \eta^{\mu\nu} -2\eta_r\left(\sigma^{\mu\nu} -\tau_\pi \chi^{\mu\nu}-\tau_\omega T_{2a}^{\mu\nu}\right)+\xi_\sigma T_{2b}^{\mu\nu}+\xi_\omega T_{2f}^{\mu\nu}.\, \, \, \, 
\end{eqnarray}
The first order term $\sigma_{\mu\nu}$ is called shear stress and is given by
\begin{eqnarray}\label{sigmadef}
 \sigma^{\mu\nu} &=& \frac{1}{2} P^{\mu\alpha}P^{\nu\beta}\left(\partial_\alpha u_\beta+\partial_\beta u_\alpha-\frac{2}{d+1}\partial.u\right).
\end{eqnarray}
The transport $\eta_r$ appearing at the first order is the shear viscosity coefficient. Since the fluid under consideration has scale invariance, the bulk viscosity coefficient is zero. 

The second order terms are given by, 
\begin{equation}
\label{secondstresscomp}
\begin{split}
T^{\mu\nu}_{2a} &=-\left({\omega^\mu}_\lambda \sigma^{\lambda\nu}+{\omega^\nu}_\lambda \sigma^{\lambda\mu}\right),
\quad 
T^{\mu \nu}_{2b} ={ \sigma^\mu}_ \alpha{\sigma^{\nu}}_{\alpha}-\frac{1}{3}P^{\mu \nu}\sigma^{\alpha\beta}\sigma_{\alpha\beta}, \\
T^{\mu\nu}_{2c} &=( \partial.u) \sigma^{\mu\nu}, \quad T^{\mu\nu}_{2d} =(u.\partial u^\mu) (u.\partial u^\nu) -\frac{1}{3}P^{\mu\nu} (u.\partial u^\alpha) (u.\partial u_\alpha),\\
T^{\mu\nu}_{2e} &= \frac{1}{2}P^{\mu\alpha}P^{\nu\beta} u^\gamma\partial_\gamma(\partial_\alpha u_\beta+\partial_\beta u_\alpha)-\frac{P^{\mu\nu}}{3} P^{\alpha\beta}(u.\partial)(\partial_\alpha u_\beta),\\
T^{\mu\nu}_{2f} &= {\omega^\mu}_\lambda \omega^{\lambda\nu}+\frac{P^{\mu\nu}}{3}\omega^{\alpha\beta}\omega_{\alpha\beta},\\
 \chi^{\mu\nu} &= \frac{1}{3}T_{2c}^{\mu\nu}+T_{2d}^{\mu\nu}+T_{2e}^{\mu\nu}.
\end{split}
\end{equation}
The vorticity tensor $\omega^{\mu\nu}$ is 
\begin{eqnarray}
\omega^{\mu\nu} &=& \frac{1}{2}\left(\partial^\mu u^\nu-\partial^\nu u^\mu +u^\mu (u.\partial)u^\nu-u^\nu (u.\partial)u^\mu\right),
\end{eqnarray}
the projector $P^{\mu\nu}$ is given by
\begin{eqnarray}
P^{\mu\nu} &=& \eta^{\mu\nu}+u^\mu u^\nu.
\end{eqnarray}

We start with this stress tensor and reduce it over the light-cone to obtain the constitutive relations of a non-relativistic fluid living in $2+1$ dimensions.

 
\section{Non-relativistic fluid}\label{nonrel-fluid}

We now briefly review the non-relativistic fluid living in flat $2+1$ dimensions. The dynamics of non-relativistic uncharged fluid is governed by the following constitutive equations  \cite{Landaubook}:
\begin{itemize}
    \item Continuity equation : 
     \begin{equation}\label{cont}
  \partial_t \rho +\partial_i (\rho v^i) =0,
  \end{equation}
  where $\rho$ is the mass density and $v^i$ is the fluid velocity.
  \item Conservation of momentum :
  \begin{equation}\label{NS}
  \partial_t ( \rho v^i) +\partial_j t^{ij} =0,
  \end{equation}
  where $t^{ij}$ is the stress tensor.
  \item  Conservation of Energy :
  \begin{equation}\label{energycurrent}
  \partial_t (\varepsilon+\frac{1}{2} \rho v^2) +\partial_i j^{i} =0 ,
  \end{equation}
  where $j^{i}$ is the energy current and $\varepsilon$ is the energy density.
\end{itemize}
These are total \emph{four} equations. There are \emph{six} non-relativistic fluid variables in $2+1$ dimensions : two components of velocity vector $v^i$, energy density $\varepsilon$, mass density $\rho$, temperature $\vartheta$ and pressure $p$. We again consider that the fluid is in local thermal equilibrium and thermodynamic variables obey Euler relation and equation of state locally. Thus we are left with four independent fluid variables. One may consider them to be fluid velocities, temperature and pressure. 

We shall later see that non-relativistic fluid obtained under light-cone reduction is a restricted class of fluid \cite{Banerjee:2014mka,Dutta:2018xtr}. Starting  from  the relativistic  thermodynamic  relations $dE=TdS$ (first law) and Euler relation $E+P=ST$ one obtains the usual non-relativistic thermodynamics $d\varepsilon = \vartheta ds +\rho_m d\rho$ (first law) and Euler  equation $\varepsilon +p = s\vartheta+ \rho_m \rho$ if $\varepsilon + p +\rho_m\rho=  0$. Here $\rho_m$ is the mass chemical potential. The last equation puts a further  restriction on the system. Using this equation one can also consider the independent variables to be velocities, temperature and mass chemical potential.
 
At ideal order, when there is no dissipation, the stress tensor and the energy current do not contain any derivatives and are solely expressed in terms of thermodynamic quantities and velocities. In presence of dissipation the stress tensor and energy current receive corrections proportional to the first derivative of velocities and temperature. Stress tensor and energy current of a compressible, uncharged, non-relativistic fluid up to the first order in derivative expansion are given by \cite{Landaubook},
\begin{equation}
\begin{split}
    t^{ij} &= \rho v^i v^j+ p \eta^{ij} -n \tilde{\sigma}^{ij} ,\\
 j^i &= \left(\varepsilon+p +\frac{1}{2}\rho {v}^2\right)v^i + n v_j \tilde{\sigma}^{ij}-\kappa \partial^i \vartheta,
\end{split}
  \end{equation}
where $\tilde \sigma_{ij}$ is the dissipative stress tensor, is given by 
 \begin{eqnarray}
  \tilde{\sigma}^{ij} =\partial^i v^j+\partial^j v^i-\delta^{ij}\partial_kv^k.
 \end{eqnarray}  
$n$ is the shear viscosity coefficient, $\kappa$ is the thermal conductivity. Since $\tilde\sigma_{ij}$ is proportional to derivative of velocity, the first order fluid is called Newtonian fluid.

 \subsection{Second grade non-Newtonian fluid}

There are several models for non-Newtonian fluids, both empirical and theoretical. The stress tensor of the simplest, second order, incompressible, non-Newtonian fluid is given by Coleman and Noll \cite{Colenoll},
       \begin{equation}\label{RECN}
       t_{ij}=- p \delta_{ij} +n \mathcal{A}_{ij}+\alpha_1 \mathcal{B}_{ij}+\alpha_2 \mathcal{A}^{ik} {\mathcal{A}_k}^{j} ,
       \end{equation}
       where  $\mathcal{A}$ and $\mathcal{B}$ are first and second Rivlin-Ericksen tensors \cite{RE} and are given by
\begin{eqnarray}\label{rivlintensors}
\begin{split}
      \mathcal{A}^{ij} &= \partial^i v^j + \partial^j v^i, \\
     \mathcal{B}^{ij} &=  \partial^i a^j + \partial^j a^i + 2\partial^i v^k \partial^j v_k,
\end{split}
\end{eqnarray}
and  
\begin{equation}\label{accn}
  a_i =  \frac{\partial v^i}{\partial t}+v^j\partial_j v^i,
\end{equation}
is the non-relativistic acceleration. $n$ is the shear viscosity, $\alpha_1  $ and $\alpha_2$ are the first and second normal stress coefficients respectively. It was shown in \cite{DFS} that a RE fluid in thermal equilibrium satisfying Clausius-Duhem inequality has the following constraints on transports
       \begin{equation}\label{Fosdick}
       n, \alpha_1\geq 0, \quad \text{and} \quad \alpha_1+\alpha_2=0.
       \end{equation}
The first condition ensures the positivity of local entropy production. Experimental results, however, show that there exist fluids which obey \eqref{RECN}, but not the second relation in \eqref{Fosdick}. It was further shown in \cite{RajaFS} that stable fluids obey $\alpha_1 \geq 0$ but the addition of $\alpha_1$ and $\alpha_2$ may not be zero always.

Another model of differential, second grade fluid is given by Huilgol \cite{Huilgol}. In our notation the corresponding stress tensor is
       \begin{equation}\label{huilgol}
        t_{ij}=- p \delta_{ij}+n \mathcal{A}_{ij} +{\alpha_1'} \left(\mathcal{B}_{ij}-\left(\tilde{\omega}^{ik}{\mathcal{A}_{k}}^j-\mathcal{A}^{ik}{{\tilde{\omega}}_k}^j\right)\right)+{\alpha_2} \mathcal{A}^{ik} {\mathcal{A}_k}^{j} ,
       \end{equation} 
where 
\begin{equation}\label{vorticity}
    \tilde{\omega}^{ik}=\partial^i v^k-\partial^k v^i,
\end{equation}
is the non-relativistic vorticity tensor. In the subsequent section we shall see that these two models of non-Newtonian fluid are subclass of a more generic second order non-relativistic fluid obtained from its relativistic counterpart under light-cone reduction.

\section{Light-cone reduction of a second order relativistic fluid}\label{L-C-R}

Light-cone reduction of relativistic conservation equation \eqref{rel:const} works in the following way. We first write down the metric of $3+1$ dimensional flat spacetime as $ds^2=-2dx^+dx^-+\Sigma_{i=1}^2 (dx^i)^2$, where $x^\pm$ are light-cone directions. We then take $x^-$ direction to be an isometry direction, \emph{i.e.} only those solutions of the relativistic energy-momentum tensor, that do not depend on $x^-$ are considered. Next, we identify $x^+$ direction with the non-relativistic time. Expressing different components of \eqref{rel:const} in light-cone coordinates and comparing them with \eqref{cont}, \eqref{NS} and \eqref{energycurrent} we find the relations between non-relativistic quantities and components of relativistic energy momentum tensor, valid at each order of derivative expansion \cite{Rangamani:2008gi,Rangamani:2009zz}. The relations are given by
    \begin{equation} \label{rel-nonrel}
    \begin{split}
        T^{++} &= \rho , \quad T^{+i} = \rho v^i , \quad T^{+-} = \varepsilon+\frac{1}{2}\rho v^2 , \\
 & \hspace{1cm} T^{-i} = j^i, \quad
 T^{ij} = {t}^{ij}.
    \end{split}
    \end{equation}
Using these, different non-relativistic constitutive relations were computed in \cite{Rangamani:2008gi,Brattan:2010bw,Banerjee:2014mka} from relativistic fluid up to first order in derivative expansion. We shall follow the similar procedure to write down the second order non-relativistic constitutive relations.

\subsection{Second order fluid}
  
Using the above reduction prescription we compute non-relativistic mass density, energy density, velocities, pressure, stress tensor and energy current up to second order. Expressions are quite cumbersome. We present the expressions of the stress tensor in the main text. Other expressions (mass density, pressure, energy density, energy current, velocities and mass chemical potential) have been provided in the appendix. 

Non-relativistic stress\footnote{To derive the non-relativistic stress tensor we have used relativistic constraint equations \eqref{secondconst}, and the identities $\frac{\eta_r}{{u^+}^2}\left(\partial^i u^+-\frac{u^+\partial^i P}{E+P}\right)=-\kappa\partial^i\vartheta$ and  $E+P=\vartheta s$.} tensor up to second order in derivative expansion is given by,
\begin{eqnarray}\label{finalsecondtij}
\begin{split}
t^{ij} &= p\delta^{ij} + \rho v^i v^j-n\tilde{\sigma}^{ij}+\tilde{\xi}_\sigma\tilde{\sigma}^{ik}{\tilde{\sigma}_k}^{j}+\tilde{\tau} _\pi\mathcal{B}^{ij} +\tilde{\tau}_\omega\left(\tilde{\omega}^{ik}\tilde{\sigma}_k^{j}-\tilde{\sigma}^{ik}\tilde{\omega}_k^j\right) +\tilde{\xi}_\omega \tilde{\omega}^{ik}\tilde{\omega}_k^j\\
&-\frac{\kappa^2\rho}{\vartheta s}\left(\frac{1}{\vartheta s}+\frac{1}{n^2}\left(\tilde{\xi}_\omega-\tilde{\xi}_\sigma-2\left(\tilde{\tau}_\pi+\tilde{\tau}_\omega\right)\right)\right)\partial^i\vartheta\partial^j \vartheta
+\frac{n\kappa}{\vartheta s}\left(\partial^i\partial^j\vartheta+\partial^j\partial^i\vartheta\right)\\
&-4{\tilde{\xi}_\omega}\frac{\rho}{\vartheta s}a^i a^j+\frac{2\kappa\rho}{\vartheta s}\left(\frac{\tilde{\tau}_w}{n}-\frac{\tilde{\xi}_\omega}{n}+\frac{2n}{\vartheta s}\right)\left(a^i \partial^j\vartheta+a^j \partial^j\vartheta\right),
\end{split}
\end{eqnarray}
where $a^i$, $\tilde{\omega}^{ij}$ are non-relativistic acceleration and vorticity tensor given in by equation \eqref{accn} and \eqref{vorticity} respectively. $\mathcal{B}^{ij}$ is second Rivlin-Ericksen tensor mentioned as before (equation \eqref{rivlintensors}). The expression for $\rho$, $p$ and $v^i$ are given by \eqref{nondensity}, \eqref{nonpressure} and \eqref{non-vel} respectively in appendix. Non-relativistic temperature $\vartheta$ is given by
\begin{equation}
    \vartheta = \frac{T}{u^+},
\end{equation}
and $s=S u^+$ is non-relativistic entropy density at the leading order. Non-relativistic transports $n$, $\kappa$, $\tilde\xi_\sigma$, $\tilde \xi_\omega$, $\tilde \tau_\pi$ and $\tilde \tau_\omega$  are determined in terms of relativistic transports and other variables. The relations are given in table \ref{table}.
      \begin{center}
      \begin{table}[h]\label{table}
      \centering 
       \begin{tabular}{ |c|c|c| }
    \hline
    Rel. transport coefficient & Non-rel. transport coefficient\\
    \hline
     $\eta_r$ & $n=\eta_ru^+ \left(1+\tau_\pi {u^+}(\partial.v)\right)$\\
        \hline 
      & $\kappa = \frac{\eta_r}{T}$\\
        \hline   
    ${\xi}_\sigma$ & $\tilde{\xi}_\sigma =\frac{ {u^+}^2}{4}\left(\xi_\sigma-4\eta_r \tau_\pi\right)$\\
        \hline
           $\xi_\omega $ & $\tilde{\xi}_\omega =\frac{ {u^+}^2}{4}\xi_\omega$\\
        \hline
       $\tau_\omega$ &  $\tilde{ \tau}_\omega =-\frac{ {u^+}^2}{2}\eta_r\left(\tau_\pi+\tau_\omega\right)$\\
            \hline
        $\tau_\pi$ & $\tilde{\tau}_\pi =\eta_r {u^+}^2\tau_\pi $ \\
        \hline
     \end{tabular}
     \caption{Relation among transport coefficients of relativistic and non-relativistic fluid.}
     \label{table}
     \end{table}
 \end{center}

The first four second derivative terms in non-relativistic stress tensor \eqref{finalsecondtij} are functions of derivatives of velocities. These four terms come under direct light-cone reduction of relativistic terms $\chi^{\mu\nu}$, $T_{2a}^{\mu\nu}$, $T_{2b}^{\mu\nu}$ and $T_{2f}^{\mu\nu}$ terms in \eqref{secondstress}. Among these four terms the first and the second term are same as the second and the first Rivlin-Ericksen terms\footnote{However, the light-cone reduced non-relativistic fluid that we are considering is \emph{not} incompressible.} respectively with $\tilde \xi_\sigma= \alpha_2$ and $\tilde\tau_\pi = \alpha_1$. Thus, RE fluid is a subclass of \eqref{finalsecondtij}. The Huilgol model \eqref{huilgol} contains another term proportional to the product of vorticity and viscous stress tensor, is also present in \eqref{finalsecondtij} but comes with an independent transport $\tilde\tau_\omega$. Therefore, Huilgol fluid is another subclass of \eqref{finalsecondtij} with $\tilde\tau_\omega = -\tilde\tau_\pi=-\alpha_1'$. As mentioned in section \ref{nonrel-fluid} that non-relativistic fluid has two independent thermodynamic variables, constitutive relations can depend on the derivatives of these two thermodynamic variables and velocities. Our light-cone reduced non-relativistic fluid also contains these terms.

From table \ref{table} we see that the second ordered  non-relativistic transport coefficients satisfy the following relation
\begin{equation}\label{nonrelation}
\tilde{\xi}_\sigma+\tilde{\tau}_\omega+\tilde{\tau}_\pi=\frac{{u^+}^2}{4}\left({\xi}_\sigma-2\eta_r\left(\tau_\pi+\tau_\omega\right)\right).
\end{equation}

\subsubsection{Holographic non-relativistic fluid}

Using the AdS/CFT correspondence Policastro, Son and Starinets computed the holographic value of shear viscosity coefficient for a conformal fluid and showed that the ratio shear viscosity to entropy density for a holographic fluid is universal \cite{Policastro:2001yc}. Holographic values of second order transports were first computed by \cite{Bhattacharyya:2008jc,Baier:2007ix,Natsuume:2007ty} independently. Universality of second order relativistic transports of conformal, non-conformal holographic fluid, fluid with higher derivative gravity dual have been discussed in \cite{Bhattacharyya:2008mz,Haack:2008xx,Shaverin:2012kv,Grozdanov:2014kva,Kleinert:2016nav,Grozdanov:2016fkt}.

The holographic values of non-relativistic transports can be computed in two different ways. One can use the dictionary between the non-relativistic and relativistic transports (table \ref{table}) and then use the holographic values of relativistic transports computed in \cite{Bhattacharyya:2008jc,Baier:2007ix,Natsuume:2007ty}. In the second approach, one constructs the holographic dual of a Schr\"odinger fluid using the TsT transformations and compute the transports by writing down the boundary stress tensor. These two methods are equivalent up to first order in derivative expansion \cite{Dutta:2018xtr}. The first order non-relativistic transports were computed in \cite{Rangamani:2008gi,Brattan:2010bw} using the first method. Here we follow the same method to compute the holographic values of second order transports.

It was shown in \cite{Bhattacharyya:2008mz,Haack:2008xx} that in any arbitrary dimensions $d>2$, uncharged relativistic conformal fluids that admit a gravity dual respect a universal relation among the second order transport coefficients,
\begin{equation}\label{reltrans}
\xi_\sigma =2\eta_r(\tau_\omega+\tau_\pi).
\end{equation}
This relation can easily be checked by computing the transports in $AdS$-Schwarzschild background \cite{,Bhattacharyya:2008mz,Bhattacharyya:2008jc,Baier:2007ix}
\begin{eqnarray}
\xi_\sigma=\frac{4\eta_r}{2\pi T},\quad \tau_\pi =\frac{2-\log{2}}{2\pi T}, \quad  \tau_\omega =\frac{\log{2}}{2\pi T} \quad \text{and}\quad \xi_\omega=0.
\end{eqnarray}
Holographic constraint on relativistic transports poses restriction on non-relativistic transport that stemmed from a holographic relativistic fluid. Using \eqref{nonrelation}, we find that the second order non-relativistic transports satisfy
\begin{equation}\label{nonrelation1}
        \tilde{\xi}_\sigma+\tilde{\tau}_\omega+\tilde{\tau}_\pi=0.
        \end{equation}
On this constraint the stress tensor \eqref{finalsecondtij} is given by
     \begin{eqnarray}\label{holotij}
     \begin{split}
       t^{ij} &= \rho v^i v^j-n\tilde{\sigma}^{ij}+p\delta^{ij}+\tilde{\xi}_\sigma\tilde{\sigma}^{ik}\tilde{\sigma}_k^{j}+\tilde{\tau} _\pi\mathcal{B}^{ij} +\tilde{\tau}_\omega\left(\tilde{\omega}^{ik}\tilde{\sigma}_k^{j}-\tilde{\sigma}^{ik}\tilde{\omega}_k^j\right) +\frac{2 n\kappa}{\vartheta s}\partial^i\partial^j\vartheta\\
      &-\frac{\kappa^2\rho}{\vartheta s}\left(\frac{1}{\vartheta s}+\frac{\tilde{\xi}_\sigma}{n^2}\right)\partial^i\vartheta\partial^j \vartheta
       +\frac{2\kappa\rho}{\vartheta s}\left(\frac{\tilde{\tau}_w}{n}+\frac{2n}{\vartheta s}\right)\left(a^i \partial^j\vartheta+a^j \partial^j\vartheta\right).
     \end{split}
      \end{eqnarray}
Comparing \eqref{holotij} with the stress tensor of RE fluid \eqref{RECN} we find
       \begin{equation}
    \alpha_1=\tilde{ \tau}_\pi = \frac{n}{2\pi\vartheta}\left(2-\log{2}\right),
       \quad \alpha_2 =\tilde{\xi}_\sigma = \frac{n}{2\pi\vartheta}\left(\log{2}-1\right).
       \end{equation}
Holographic values of other second order non-relativistic transports are given by
   \begin{eqnarray}
    \tilde{\tau}_\omega=-\frac{n}{2\pi\vartheta} \quad \text{ and}\quad \tilde{\xi}_\omega=0.  
   \end{eqnarray}
The stress tensor \eqref{holotij} is a generalised version of Huilgol stress tensor  \eqref{huilgol} with $\tilde{\tau}_\omega=-\tilde{\tau}_\pi$. The constraint relation \eqref{nonrelation1} is the generalization of the relation given by Dunn and Fosdick \eqref{Fosdick}. Apart from the regular velocity dependent terms, our expression \eqref{holotij} also contains terms dependent on derivatives of temperature. Thus, we have constructed the most generic holographic non-relativistic stress tensor corrected up to second order in the derivative expansion.

As mentioned in section \ref{nonrel-fluid}, the light-cone reduced non-relativistic fluid belongs to a restricted class where the mass chemical potential $\rho_m$ follows the relation $\varepsilon+p+\rho_m \rho =0$. One can use this relation to find the mass chemical potential corrected up to second order \eqref{masschempot}. $\rho_m$ does not receive any correction at first order as before \cite{Banerjee:2014mka}, however, it has corrections at the second order.

\section{Conclusion and Outlook}\label{Diss}

In this paper, we have studied the holographic constraint on second order transports of an uncharged non-relativistic fluid obtained from a relativistic uncharged fluid with Weyl invariance by light-cone reduction. The resultant holographic stress tensor has seven terms at the second order with four transports. We also compute the holographic values of these second order transports and observe that three of them satisfy a universal relation : \emph{sum of the three coefficients are zero}. The expression of the non-relativistic stress tensor obtained from its relativistic counterpart is a generalisation of two well-known Rheological models, namely Rivlin-Ericksen and Huilgol model.
  
Our results open up the platform to study the second order stress tensor for a non-relativistic charged fluid under light-cone reduction of charged, second order relativistic fluid. The construction of second order constitutive relations of the non-relativistic fluids from a non-relativistic gravity dual is another important problem to look at. Second order entropy current for uncharged/charged non-relativistic fluid can also be studied. It would be interesting to find the constraints on second order non-relativistic transports imposed by the second law of thermodynamics.

\section*{Acknowledgement}
We would like to thank Nabamita Banerjee, Sayantani Bhattacharyya and Debangshu Mukherjee for insightful discussions. The work of SD is supported by the grant no. EMR/2016/006294 and 
MTR/2019/000390 from the SERB, Government of India. SD also acknowledges the Simons Associateship of the Abdus Salam ICTP, Trieste, Italy. Finally, we are grateful to people of India for their
unconditional support towards researches in basic sciences.  

\appendix

\section{Detailed calculation of non-relativistic quantities} 

The first order relativistic fluid satisfies the following relations
\begin{eqnarray}\label{secondconst}
\begin{split}
(u.\partial)E+(E+P) (\partial.u)  &=\eta_r \left(\partial_\mu u_\nu+\partial_\nu u_\mu\right) \sigma^{\mu\nu},\\
P^{\mu\nu}\partial_\mu P +(E+P)(u.\partial)u^\nu 
&=2 \eta_r\left(\partial_\mu \sigma^{\mu\nu}-u^\nu \sigma_{\alpha\beta}\sigma^{\alpha\beta}\right).
\end{split}
 \end{eqnarray}
The relativistic fluid velocity is normalised : $u^\mu u_\mu = -1$. We use this relation to replace $u^-$ component in terms of other components 
\begin{eqnarray}
 & u^- & = \frac{1}{2u^+}\left( 1+ u_k u^k\right)\label{uminus}.
\end{eqnarray}
 
Computing the $++$ component of the relativistic energy-momentum tensor \eqref{secondstress} and identifying that with the non-relativistic mass density $\rho={T^{++}}$,  we obtain
\begin{eqnarray}\label{nondensity}
&&\rho = (E+P)(u^+)^2-\frac{2\eta_r^2 u^+}{E+P}\partial_k \mathcal{Y}^k-\frac{{u^+}^2}{3}\left(\xi_\sigma-\frac{16\eta_r^2}{E+P}\right)\sigma_{\alpha\beta}\sigma^{\alpha\beta}+\frac{\xi_\omega}{3}{u^+}^2\omega_{\alpha\beta}\omega^{\alpha\beta}\nonumber \\&&+\left(2\eta_r \left(\tau_\pi+\tau_\omega\right)-\xi_\omega\right)  \frac{u^+\mathcal{Y}_k \partial^k P}{E+P}
+\left(4\eta_r \tau_\omega +\xi_\sigma-\xi_\omega\right) \frac{\mathcal{Y}_k \mathcal{Y}^k}{4} -\xi_\omega \frac{{u^+}^2\partial_k P\partial^k P}{(E+P)^2},
\end{eqnarray}
where 
\begin{equation}
   \mathcal{Y}^\alpha=\mathcal{Y}^{\alpha+}=\partial^\alpha u^+ + u^+ (u.\partial)u^\alpha-\frac{u^+u^\alpha}{3}(\partial.u) ,
\end{equation}
and
\begin{eqnarray}
\mathcal{Y}^{\mu\nu}=\partial^\mu u^\nu+u^\nu u^\gamma \partial_\gamma u^\mu-\frac{1}{3}u^\mu u^\nu (\partial.u).
\end{eqnarray}
At first order it reduces to
 $\mathcal{Y}^\alpha=\left(\partial^\alpha u^+ - \frac{u^+ \partial^\alpha P}{E+P}\right)$.
As before the non-relativistic mass density does not receive any correction at first order for the conformal fluid \cite{Rangamani:2008gi,Banerjee:2008th}.
   
We now use the mapping $T^{+i} =\rho v^i$, to compute non-relativistic velocity in terms of relativistic data. Computing $+i$ component of $T^{\mu\nu}$ we find
\begin{eqnarray}\label{non-vel}
&& v ^i =  \frac{u^i}{u^+}-\frac{\eta_r}{\rho}\mathcal{Y}^i +\frac{\eta_r^2{u^+}^2}{\rho^2}\left(\partial_k \mathcal{Y}^k-\frac{8}{3}u^+\sigma_{\alpha\beta}\sigma^{\alpha\beta}\right)u^i+\frac{\eta_r}{\rho}\tau_\omega\Big(\Big(\mathcal{Y}^k+\frac{u^+\partial^k P}{E+P}\Big)\Big(\partial_k u^i\nonumber\\&&-\frac{u^i}{u^+}\mathcal{Y}_k-\frac{u^i\partial_k P}{E+P}\Big)+\Big(u.\partial P-u^k\partial_k P\Big)\frac{\mathcal{Y}^i}{E+P}+\frac{u^+}{E+P}\Big(2u.\partial P-u^k\partial_k P\Big)\frac{\partial^i P}{E+P}\nonumber\\ 
 &&+u^+\frac{\partial^i u^k\partial_k P}{E+P}\Big)+\frac{\xi_\sigma}{4\rho}\Big(\mathcal{Y}_+ \mathcal{Y}^i+\frac{4u.\partial P}{E+P}\mathcal{Y}^i+\mathcal{Y}_k\Big(\partial^k u^
i+\partial^i u^k-\frac{u^k\partial^i P}{E+P} \Big)\nonumber\\ 
 &&-\mathcal{Y}_k\Big(\mathcal{Y}^k+\frac{u^+\partial^k P}{E+P}\Big)\frac{u^i}{u^+}\Big)+\frac{\xi_\omega}{4\rho}\Big(\Big(\mathcal{Y}_+ +2\frac{u^+\partial_+P}{E+P}\Big)\mathcal{Y}^i-\Big(\mathcal{Y}^k+2\frac{u^+\partial^k P}{E+P}\Big)\Big(\partial_k u^i\nonumber\\ 
 &&-\partial^i u_k\Big)+\frac{\partial^i P}{E+P}\Big(u_k \mathcal{Y}^k  +2u^+\mathcal{Y}_+ +\frac{2u^+}{E+P}\Big(2 u.\partial P-u^k\partial_k P\Big)\Big)+\Big(\mathcal{Y}^k\mathcal{Y}_k\nonumber\\ 
 &&+3u^+\frac{\mathcal{Y}^k\partial_k P}{E+P}+2{u^+}^2\frac{\partial_k P\partial^k P }{(E+P)^2}\Big)\frac{u^i}{u^+}\Big)+\frac{\eta_r}{\rho}\tau_\pi\Big(u.\partial \mathcal{Y}^i+\frac{u^+\partial^k P}{E+P}\left(\partial^k u^i-\frac{u^i\partial^k P}{E+P}\right)\nonumber\\ 
 &&-\frac{u.\partial P}{E+P}\Big(\mathcal{Y}^i-\frac{u^+\partial^i P}{E+P}\Big)-u^i \frac{\mathcal{Y}_k\partial^k P}{E+P}
+u^+\frac{\partial^\alpha P\partial^i u_\alpha}{E+P}\Big).
\end{eqnarray}

Second order corrected non-relativistic equilibrium pressure in terms of relativistic variables can be obtained from the diagonal terms in $T^{ij}$ and is given by,
\begin{eqnarray}\label{nonpressure}
p &=& P+\frac{\eta_r^2}{\rho}\Big(2 u^+ \partial_k \mathcal{Y}^k-\frac{8}{3}{u^+}^2\sigma_{\alpha\beta}\sigma^{\alpha\beta}-2\mathcal{Y}_k\mathcal{Y}^k+6 {u^+}^3 a_k\mathcal{Y}^k\Big)\nonumber\\&&-\eta_r \tau_\pi {u^+}^2\left(\partial.a-\partial_k v^m\partial_m v^k+(\partial.v)^2\right)-\frac{\xi_\sigma}{3}\sigma_{\alpha\beta}\sigma^{\alpha\beta}{+\frac{\xi_\omega}{3}\omega_{\alpha\beta}\omega^{\alpha\beta}}.
\end{eqnarray}
  
Computation of $+-$ component of the relativistic stress tensor and  the identification $T^{+-}=\varepsilon+\frac{1}{2}\rho v^2$,  give us the non-relativistic energy density
\begin{eqnarray}\label{nonen}
  &&    \varepsilon=   \frac{E-P}{2}- \frac{1}{2{u^+}^2}\left( \frac{{\eta_r}^2 }{(E+P)}+\eta_r \tau_\omega+\frac{\xi_\sigma-\xi_\omega}{4}\right)\mathcal{Y}_k \mathcal{Y}^k-\frac{1}{2{u^+}^2}\Big[-2\frac{\eta_r^2 u^+}{E+P}\partial_k\mathcal{Y}^k\nonumber\\
      &&+\Big(2\eta_r \left(\tau_\pi+\tau_\omega\right)-\xi_\omega\Big) u^+ \frac{\mathcal{Y}_k \partial^k P}{E+P}
 -\frac{{u^+}^2}{3}\left(\xi_\sigma-\frac{16\eta_r^2}{E+P}\right)\sigma_{\alpha\beta}\sigma^{\alpha\beta}+\frac{\xi_\omega}{3}{u^+}^2\omega_{\alpha\beta}\omega^{\alpha\beta}\nonumber\\&&-\xi_\omega {u^+}^2\frac{\partial_k P\partial^k P}{(E+P)^2}\Big].
      \end{eqnarray}

Non-relativistic energy current density can be computed by identifying $T^{-i}=j^i$. It is given by
\begin{eqnarray}\label{finalcurrent}
&& j^{i} =   \left(\varepsilon+p+\frac{1}{2}\rho v^2\right)v^i-n v_k \tilde{\sigma}^{ki}  -\kappa\left(1+\frac{3n}{2\vartheta s}(\partial.v)\right)\partial^i\vartheta+\frac{\kappa}{n}\left(\tilde{\xi}_\sigma-\tilde{\tau}_\omega\right)\partial_k\vartheta \tilde{\sigma}^{ki} \nonumber\\&&
-\left(\frac{n^2}{\vartheta s}+2\tilde{\tau}_\omega\right) a_k \tilde{\sigma}^{ki} +
\frac{n\kappa}{\vartheta s}v_k\left(\partial^k\partial^i\vartheta+\partial^i\partial^k\vartheta\right)-\frac{\kappa }{n}\left(\frac{n^2}{\vartheta s}+\tilde{\xi}_\omega+\tilde{\tau}_\omega\right)\partial_k\vartheta\omega^{ki}\nonumber\\&&+\tilde{\xi}_\omega v_k \tilde{\omega}^{kl}\tilde{\omega}_l^i
-\frac{\kappa^2\rho}{\vartheta s}\left(\frac{1}{\vartheta s}+\frac{\tilde{\xi}_\omega-\tilde{\xi}_\sigma-2\left(\tilde{\tau}_\pi+\tilde{\tau}_\omega\right)}{n^2}\right)v_k\partial^k\vartheta\partial^i\vartheta
-4\tilde{\xi}_\omega\frac{\rho}{\vartheta s}a_k v^k a^i\nonumber\\&&+\tilde{\tau}_\pi v_k \mathcal{B}^{ki} +\tilde{\xi}_\sigma v_k \tilde{\sigma}^{kl}{ \tilde{\sigma}_l}^{i}+\frac{2\kappa\rho}{n\vartheta s}\left(\frac{2n^2}{\vartheta s}+\tilde{\tau}_\omega-\tilde{\xi}_\omega\right)v_k \left(\partial^k\vartheta a^i +\partial^i\vartheta a^k\right)-2\tilde{\xi}_\omega a_k\tilde{\omega}^{ki} \nonumber\\
&&+\tilde{\tau}_\omega v_k\left(\omega^{kl}\tilde{\sigma}_l^i-\tilde{\sigma}^{kl}\omega_l^i\right)-\left(\tilde{\tau}_\pi-\frac{n^2}{\vartheta s}\right)\left(\frac{\vartheta s}{\rho}\right)\partial^i(\partial.v)+\frac{n^2}{\rho}\partial_k \tilde{\sigma}^{ki}. \end{eqnarray}

Using the relation $\varepsilon +p+\rho_m \rho=0$, we find the mass chemical potential as follows,

 \begin{eqnarray}\label{masschempot}
 && \rho_m = -\frac{1}{2{u^+}^2}+\frac{2n^2}{\vartheta^2s^2{u^+}^2}\tilde{\sigma}^{kl}\tilde{\sigma}_{kl}+\frac{\tilde{\tau}_\pi}{\vartheta s {u^+}^2}\left((\partial.v)^2-\partial_k v^m\partial_m v^k\right)+\left(\frac{4n^2}{\vartheta s}-\tilde{\tau}_\pi\right)\frac{\partial_k\partial^k \vartheta}{\vartheta^2s {u^+}^4}\nonumber\\
 &&+\left(\tilde{\tau}_\pi+2\tilde{\tau}_\omega-\tilde{\xi}_\omega+\tilde{\xi}_\sigma-\frac{7n^2}{2\vartheta s}\right)\frac{\partial_k\vartheta\partial^k \vartheta}{\vartheta^3s{u^+}^4}
 +\left(2\tilde{\tau}_\pi+4\tilde{\tau}_\omega-4\tilde{\xi}_\omega-\frac{2n^2}{\vartheta s}\right)\frac{\partial_k\vartheta\partial^k {u^+}}{\vartheta^2 s{u^+}^3}\nonumber\\
 &&-\frac{\tilde{\tau}_\pi}{\vartheta s {u^+}^5}\partial_k\partial^k u^+ + \frac{3\tilde{\tau}_\pi-4\tilde{\xi}_\omega}{\vartheta s {u^+}^6}\partial_k u^+\partial^k u^+ .
 \end{eqnarray}

\bibliographystyle{JHEP}
\bibliography{rivlin}
\end{document}